\documentclass[11pt,a4paper,bibtotoc,abstract=on]{scrreprt}

\usepackage[pdftex]{graphicx}
\usepackage[caption=false,font=footnotesize,labelfont=sf,textfont=sf]{subfig}
\usepackage{hyperref}
\usepackage{pdflscape}
\usepackage{float}
\usepackage{amsmath,array,url}
\usepackage{amssymb}
\usepackage{xcolor}
\usepackage{csquotes}
\usepackage{longtable}
\usepackage{tabularx}
\usepackage{booktabs}
\usepackage{geometry}

\usepackage[style=authoryear,natbib=true,backend=bibtex]{biblatex}
\begin{document}

\title{A Survey of Music Generation in the Context of Interaction}

\author{
Ismael Agchar$^1$ \and
Ilja Baumann$^1$ \and
Franziska Braun$^1$ \and
Paula~Andrea P\'erez-Toro$^1$ \and
Korbinian Riedhammer$^1$ \and
Sebastian Trump$^2$ \and
Martin Ullrich$^2$
\thanks{This work was supported by LEONARDO, a joint project by Technische Hochschule Nürnberg and Hochschule für Musik Nürnberg. Authors are listed in alphabetical order.}
}
\publishers{
$^1$Technische Hochschule Nürnberg\\
$^2$Hochschule für Musik Nürnberg\\[2cm]
\texttt{korbinian.riedhammer@th-nuernberg.de}\\
\texttt{sebastian.trump@hfm-nuernberg.de}
}

\newcommand{\eg}{\textit{e.g.}, }  
\newcommand{\ie}{\textit{i.e.}, }  
\newcommand{\cf}{\textit{cf.} } 

\maketitle

\chapter*{A Note to the Reader}
This report largely written back in Spring 2021, but unfortunately never submitted to peer review.
In the meantime, two extensive reviews of similar nature have been published by \citeauthor{CIVIT2022118190} (peer reviewed) and \citeauthor{zhao2022review} (arxiv).
We believe that this manuscript still adds value to the scientific community, since it focuses on music generation in an interactive, hence real-time, scenario, and how such systems can be evaluated.

\begin{abstract}
In recent years, machine learning, and in particular generative adversarial neural networks (GANs) and attention-based neural networks (transformers), have been successfully used to compose and generate music, both melodies and polyphonic pieces.
Current research focuses foremost on style replication (\eg generating a Bach-style chorale) or style transfer (\eg classical to jazz) based on large amounts of recorded or transcribed music, which in turn also allows for fairly straight-forward ``performance'' evaluation.
However, most of these models are not suitable for human-machine co-creation through live interaction, neither is clear, how such models and resulting creations would be evaluated.
This article presents a thorough review of music representation, feature analysis, heuristic algorithms, statistical and parametric modelling, and human and automatic evaluation measures, along with a discussion of which approaches and models seem most suitable for live interaction.
\end{abstract}

\tableofcontents



\chapter{Introduction}\label{sec:music_generation}

There is no single definition of music that is universally accepted across human cultures~\citep{Nettl_2014m}.
Moreover, the sound production of nonhuman animals can be classified as music, too~\citep{Slater_Doolittle_2014am}.
Nevertheless, most of the reviewed articles in this overview seem to start from a narrower definition of music, at least implicitly.
A definition based on the tradition of Western tonal music could read as follows: 
Music is a set of sounds and silences that are organized logically and coherently.
This concept of music is related to a series of attributes: sounds,  humans, and aesthetics~\citep{wallin2001tom}.
Humans generate these sounds, that can be described as music notes, through different means such as their voices or musical instruments.
The choice and order of these sounds are influenced by the predominant aesthetics of the given culture.
Music has certain properties ruled by a series of fundamental principles related to melody, harmony, rhythm, and timbre.
The art of music creation can be defined as organizing sounds in time by following these rules.
Music sounds are produced by some physical processes that have a periodic character such as a vibrating string or the air inside a wind instrument. 
This set of sounds is a collection of tones of different frequencies that can be treated as an analog signal.
Consequently, this process can also be described via mathematical models giving way to automatic music generation.
Automatic music generation is a topic-focused on the music creation starting from the general idea that statistical and learning models can learn how to generate something that did not exist before with minimal human intervention. 
This is a challenging task due to the enormous level of complexity and abstraction from the point of view of creativity, therefore it requires starting from concepts that already existed.
Nowadays, statistical and parametric models are commonly used to automatically generate music pieces. 
The statistical models in music generation aim to assign probabilities to events conditioned to earlier events in a time sequence.
There are several statistical approaches used in music such as Markov chains, restricted Boltzmann machines, finite-states, among others. 
Parametric models, especially those based on deep learning, have become one of the leading techniques for automated music generation.
The most popular deep learning networks for music generation are those based on recurrent networks and generative models.

A piece of music can be generated using one single melodic line (monophonic) or involving several sounds simultaneously (polyphonic).
On one hand, monophonic generation focuses on generating a melody using a single tune without any harmony.
Several approaches aim to generate music, extract melodic features and explore properties from monophonic melodies via mathematical models~\citep{hoover2012gcm,nakamura2016tspm,prudente2017tacm,wang2020epm}. 
On the other hand, polyphonic generation requires more complex models, since it needs to predict the probability of any combination of notes across time.
Currently, many studies~\citep{dong2018musegan,dong2018cgap,wang2020lirp,tiwari2020pmg} have been approached on the generation of music from polyphonic sounds, since it allows combining multiple melodic lines at the same time, which is a more challenging task.

There are two main challenges to focus on in music generation: (1) generate music focused on the rhythm and harmony and (2) focused on long-term structures with the ability to generate a melody.
If the generating system is to be used in an interactive setup, the algorithms or models may need to satisfy input or runtime constraints.
Generally, the rhythm and harmony is affected when considering long-term memory models, which causes that in the generation of melodies the rhythm and harmony are compromised~\citep{hebert1997rmltm}.
Some studies seek to address these two challenges.
For instance, probabilistic models are often used to synchronize rhythm and melody or model which melody notes fit given chords~\citep{roy2017svls}. 
Other kinds of approaches such as those based on recurrent networks generate the melody according to predefined rhythmic and harmony templates~\citep{de2019rcm}, while
generative models aim to compose music pieces from ``scratch''~\cite{liu2018lsgan}.

Therefore, with a focus on theoretical and technical foundations, this survey aims to highlight fundamentally different approaches to music generation and addresses the question of their potential for interactive scenarios. These are the basis of a constantly growing number of use cases, which are mentioned here, however, only in individual selected examples.

\chapter{Data and Formats}\label{sec:data_formats}


The different formats and datasets addressed in the papers reviewed in this study are presented in this section.

\section{Formats}

There are several alternatives of input formats used in automatic music generation.
However, two main types of music formats can be considered: symbolic and digital audio formats. 

\subsection{Symbolic}

This kind of representation aims to comprise logical structures based on score representations and symbolic elements, which encode musical events.
It allows generalizing concepts of music notation to model aspects of audio, music scores, or additional annotations related to a music piece. 
The digital data formats such as Musical Instrument Digital Interface (MIDI) and piano-rolls can be considered symbolic when it is based on a predefined alphabet of symbols.\\


\emph{Musical Instrument Digital Interface-MIDI:}

Nowadays, this format is one of the most used for music generation, because it ease modification and manipulation. 
Additionally, this format describes a communication protocol, which allows connecting a wide variety of electronic instruments and other digital musical tools.
A MIDI File is a series of 8-bit bytes, that follows a chunk architecture composed by a header and a track chunk~\citep{moog1986midi}. 
The header contains information related to the entire MIDI file. 
The header contains information related to the entire MIDI file as chunk type, length, format, number of track chunks, and the meaning of the delta-times.
A track chunk is simply a stream of MIDI events preceded by the delta-times.
It can represent information in up to 16 channels of a sequential stream of MIDI data, which the multiple tracks files are implemented by using several track chunks.\\

\emph{Piano-rolls:}

This format is an adaptation of the typical rolls used as mechanical methods to store piano compositions. 
For computational analyses, the piano-rolls are represented as a bi-dimensional matrix, where the vertical axis is the note pitch and the horizontal represents the time. 
An horizontal line is drawn when a note is played. the height represents which note is played.
An horizontal line is drawn when a note is played. 
The height indicate which note was played, while the onset of the note is given by the beginning of the line, the duration by the length, and the offset by the end.
The multi-track representation is composed by a set of piano-roll matrices, where each one corresponds to one specific track.\\

\emph{ABC notation:}

It is a music writing language which uses a set of ASCII characters~\citep{walshaw2000abc}. 
Although this is a language for software purposes, it is easily readable and interpretable for humans.
This notation consist of three parts: (1) the file header, carries the processing information for the entire file, (2) the tune header, contains information related to the metadata, and (3) the tune body, contains the ABC music code.
The ABC system can be used in any text editor since it is based on ASCII characters, which also it is very intuitive as the notes have a mathematical value and the syntax allows to use metadata for each tone.
Additionally, there are several programs to interpret music written in ABC notation, as well as to create a score in musical notation.

\subsection{Digital Audio}

This file format is used to store digital audio in electronic devices. 
The piece of music is recorded in digital form, where a sequence of discrete samples are taken from an analog audio waveform~\citep{zolzer2008dasp}. 
There are three main groups of formats: uncompressed, lossless compression, and lossy compression.\\

\emph{Uncompressed Formats:}

This set of formats does not use any codec or compression algorithm, allowing the audio to a close-to-exact representation of analog sound. 
they are represented by using the Pulse Code Modulation (PCM) method, which is a digital representation of raw analog audio signals and it is widely recognized as the simplest audio storage mechanism in the digital domain. 
The waveform is converted into digital bits, where the audio signal is sampled and recorded at certain pulses or intervals.
Some of the major formats in this category includes: Waveform Audio Format (WAV), Audio Interchange File Format (AIFF), and the Broadcast Wave Format (BWF).
This format is the most frequently used in automatic music generation.\\

\emph{Lossless compression Formats:}

It is a set of data compression algorithms that allows decompress back to their original size, keeping sound quality intact. The bit rates depend on the density and volume of the audio rather than on the quality.
This means that this format requires more processing time than uncompressed formats but is more space efficient compare to uncompressed formats. 
The most known lossless data compression algorithms for audio are: Free Lossless Audio Codec (FLAC) and Apple’s Lossless Audio Codec (ALAC).\\

\emph{Lossy compression Formats:}

These set of data encoding methods compress the audio by removing non-essential information, which offers the smallest file sizes. 
This format is not often used for automatic music generation because the file cannot be decompressed to its original file size, losing information. 
Additionally, the file suffers significant quality degradation, especially at higher compression rates.
Two of the most widely used formats are Mpeg Audio Layer 3 (MP3) and Advanced Audio Coding (ACC).

\section{Automated Transcription}
The concepts of harmony, rhythm and long-term structure of a piece of music are inherently symbolic.
Hence, music usually is represented symbolically when it is to be generated.
Often, some form of music analysis precedes the process of generation.
Music analysis (regarding harmony, rhythm, structure, etc.) is usually also carried out on symbolic representations.
But if the music to be analyzed is available only in digital form (as mentioned before), means of automated transcription become necessary.

Music transcription is a difficult task, even for humans from which it still demands years of practice.
\citep{benetos2018amto} give an overview of general approaches and state of the art methods for automated music transcription.



\section{Datasets}

Information about some datasets commonly used in music generation are presented in Table~\ref{tab:datasets}.
The column labeled ``References'' indicates the studies in which each dataset was used.

\newgeometry{left=0.5cm,right=0.5cm,bottom=0.5cm,top=0.5cm}
\pagestyle{empty}
\begin{landscape}
\begin{table}[htbp]
\centering
\small
\caption{Summary of datasets commonly used for automatic music generation}
\label{tab:datasets}
\begin{tabular}{lp{3cm}p{3cm}lp{10cm}}
\toprule
\textbf{Dataset} & \textbf{Description} & \textbf{Instruments} & \textbf{Format} & \textbf{References} \\ \hline
\midrule
Maestro          
	& 200 hours recorded from the International Piano-e-Competition 
	& Piano 
	& WAV/MIDI 
	& \cite{hawthorne2018efpfmd,huang2018mtlt,stgeorge2019mst,modrzejewski2019amcg,choi2020emtae} \\\hline
NSynth           
	& 305,979 musical notes from different instruments 
	& Brass, flute, guitar, keyboard, mallet, organ, reed, string, synth lead, vocal               
	& WAV/MIDI 
	& \cite{engel2017nas,andreux2018mgt,roberts2018llrgi,defossez2018sing,engel2019gansynth} \\\hline
Groove MIDI Dataset (GMD)
	& 444 hours of audio from 43 drum kits
	& Drums 
	& WAV/MIDI 
	& \cite{gillick2019lgist,callender2020ipqd,burloiu2020iled} \\\hline
J.\,S.\,Bach (JSB) Chorales
	& 100 Chorales composed by Johann Sebastian Bach
	& Voice encodings 
	& MIDI
	& \cite{hild1992harmonet,goel2014pmgrnn,chung2014egrnsm,lyu2015mhdslstm,madjiheurem2016c2v,liang2017ascbc,chi2020gmsc,huang2018mtlt} \\\hline
MusicNet
	& 330 classical music recordings
	& Piano, violin, viola, cello, clarinet, bassoon, horn, oboe, flute, harpsichord, string bass
	& WAV/MIDI
	& \cite{thickstun2016lfms,manzelli2018cdgam,manzelli2018eteamg,wang2019pnet,michelashvili2020htpag} \\\hline
Nottingham
	& 1,000 British and American folk tunes
	& --
	& ABC/MIDI
	& \cite{goel2014pmgrnn,lee2017pmggan,agarwal2018lmgrt,liang2019ms,jhamtani2019msrgan,raja2020mgtsa,cheng2020vmtgan} \\\hline
MAPS
	& 65 hours of audio recordings
	& Piano
	& WAV/MIDI
	& \cite{emiya2010maps,hawthorne2018efpfmd} \\\hline
GenImpro
	& 30 hours of free duo improvisations
	& Snaredrum, recorder, violin, double bass, percussion, trombone, cello,  soprano saxophone, flute
	& WAV/MIDI
	& \cite{trump2020evoimpro} \\\hline
MuseData
	& Music pieces from several classical music composers
	& Piano
	& MIDI
	& \cite{goel2014pmgrnn,huang2016dlm,johnson2017gpmtpn} \\\hline
Lakh
	& 174,154 multi-track piano recordings
	& Bass, drums, guitar, strings, piano
	& MIDI/Piano-roll
	& \cite{raffel2016lakh,dong2018cganpm,dong2018musegan,liang2019ms2,guan2019gansami,dong2020muspy} \\\hline
MetaMIDI (MMD)
    & 436,631 files with partial metadata
    & various
    & MIDI
    & \cite{ens2021mmd} \url{https://github.com/jeffreyjohnens/MetaMIDIDataset} \\\hline

\end{tabular}
\end{table} 
\end{landscape}
\restoregeometry
\pagestyle{plain}


Maestro, NSynth, and Groove MIDI Dataset (GMD) are created by the Magenta project. 
The Maestro dataset~\citep{hawthorne2018efpfmd} contains paired audio and MIDI piano recordings collected during 10 years in the International Piano-e-Competition. 
NSynth~\citep{engel2017nas} consists of  several musical notes with a unique pitch, timbre an envelope.
The notes were produced by using 3 sources methods (acoustic, electronic and,  synthetic) from 11 different instruments.
GMD~\citep{gillick2019lgist} considers MIDI and WAV drum recordings with annotations of velocity from human performances on a Roland TD-11 electronic drum kit.

Other datasets out of the Magenta project as JSB Chorales, MusicNet, Nottingham, MAPS, MuseData, and Lakh are considered.
JSB Chorales dataset first mentioned in~\citep{hild1992harmonet}, collects a set of single-line melodies of 100 chorales from 4 different voices. 
The voices are encoded in a series of events which form a MIDI file.
MusicNet~\citep{thickstun2016lfms} is a multi-track dataset of classical music recordings with MIDI annotated labels acquired from musical scores aligned to recordings by dynamic time warping.
Some datasets are also derived from contemporary music such as Nottingham\footnote{ifdo.ca/~seymour/nottingham/nottingham.html}, which contains 1 thousand of folks tunes.
Piano recordings are commonly used in music generation due to the variety of datasets which contains music pieces of this instrument~\citep{hawthorne2018efpfmd,hild1992harmonet,emiya2010maps}.
MAPS~\citep{emiya2010maps} is a database which contains disklavier piano recordings and synthesized audio created from MIDI files.
Another piano dataset is MuseData\footnote{http://musedata.org/}. 
It is a text-based method that consists of storing musical information allowing a score and MIDI file generation.
The GenImpro dataset \citep{trump2020evoimpro} contains 30 hours of free duo improvisations with different instruments as WAV recordings, partly transcribed in MIDI.
MuseData is derived from pieces of classical music by composers such as Beethoven, J. S. Bach, Vivaldi, among others.
Lakh dataset~\citep{raffel2016lakh} contains multi-track MIDI and piano-rolls, where some recordings have been matched and aligned to entries in the Million Song Dataset~\citep{bertin2011msd}.

\chapter{Transformed Representations}


In audio analysis the data is processed by computing compressed representations of the signal that can not capture all of the dynamic information. 
However, it is possible to obtain different representations to observe how is the behaviour in time of some aspects such as the variation of energy in the spectrum or in the chroma domain.

\section{Spectrogram}

\begin{figure}[!htpb]
    \centering
    \includegraphics[width=\linewidth]{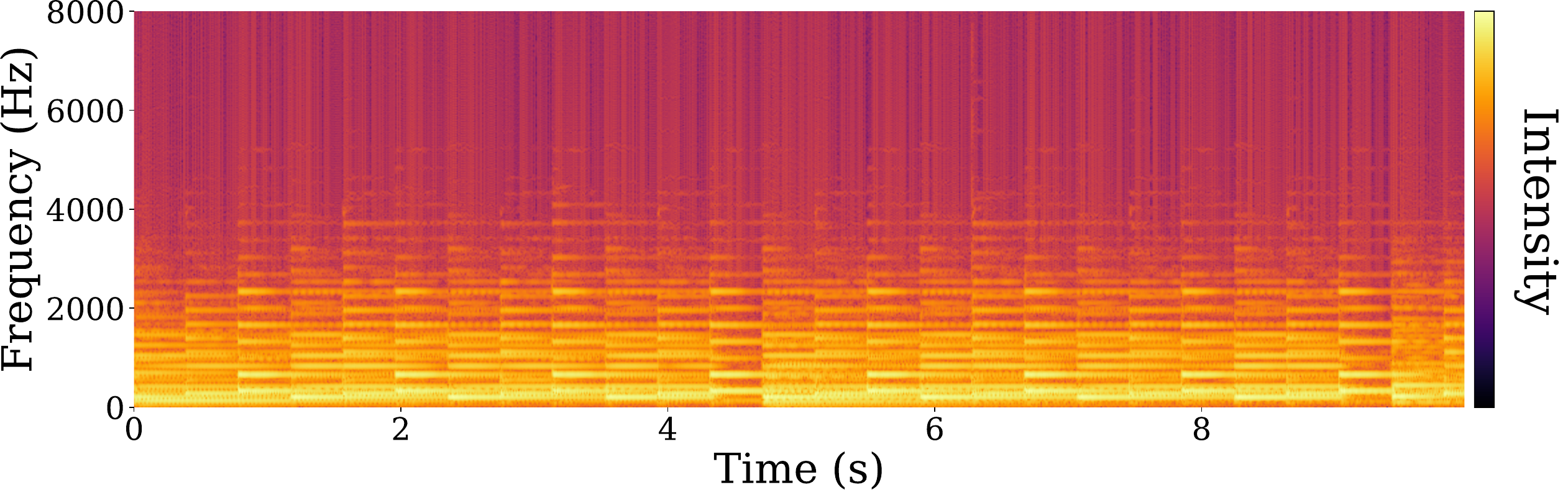}
    \caption{Log--Spectrogram representation of the Bethoveen's piano composition ``Moonlight Sonata''}
    \label{fig:spec}
\end{figure}

Commonly, in music generation the spectrum from an audio signal is used as visual representation to analyze the dynamic information embedded in the recording.
This representation allows to observe how the energy varies in the frequency domain with respect to the time. 
It can be achieved by using the Fourier transform that converts the signal into a time-frequency representation.
A variation known as the Short-Time Fourier Transform (STFT) is often used to obtain this 2-dimensional graphical representation.
The spectrogram is obtain by compute the spectral power STFT.
The temporal or frequency resolution of the analysis in the STFT depends on the window size. 
Thereby, small local translations in the time-frequency domain can be produced, which may result in smooth perceptual variations.
Furthermore, the logarithm version of theses spectrograms are highly considered since the harmonics on this scale are equidistant.
Figure~\ref{fig:spec} shows the log--spectrogram of ``Moonlight Sonata'', one of the most famous piano compositions of Bethoveen. 
For visualization purposes, the spectrogram was cut off at 8 kHz. 
The x--axis represent the time in seconds, while the y--axis represent the frequency in Hertz. 
The different color maps helps to observe how the energy intensity varies in the different frequency band. 

Consequently, several approaches considered the use of spectrograms to generate musical raw signals~\citep{engel2017nas,michelashvili2020htpag,andreux2018mgt,defossez2018sing,wang2019pnet,kalingeri2016mgdl} or even to improve the quality in the generation of symbolic music pieces\citep{manzelli2018cdgam,manzelli2018eteamg}.
Magnitude spectrograms have been widely used in music generation, since they can capture many aspects of the dynamic of a signal~\citep{engel2017nas,manzelli2018cdgam,michelashvili2020htpag,andreux2018mgt,wang2019pnet,kalingeri2016mgdl}. 
However, phase spectrograms have also been widely considered due to the fact that for some sounds their difference lies in phase changes, where  their magnitude spectrograms sometimes are very similar and hardly differentiable~\citep{engel2017nas,defossez2018sing}.
Additionally, peak frequency spectrogram are also used in few applications to quantitatively evaluate the structuring ability of the model~\citep{manzelli2018eteamg}.

\section{Mel Spectrogram}

\begin{figure}[!htpb]
    \centering
    \includegraphics[width=\linewidth]{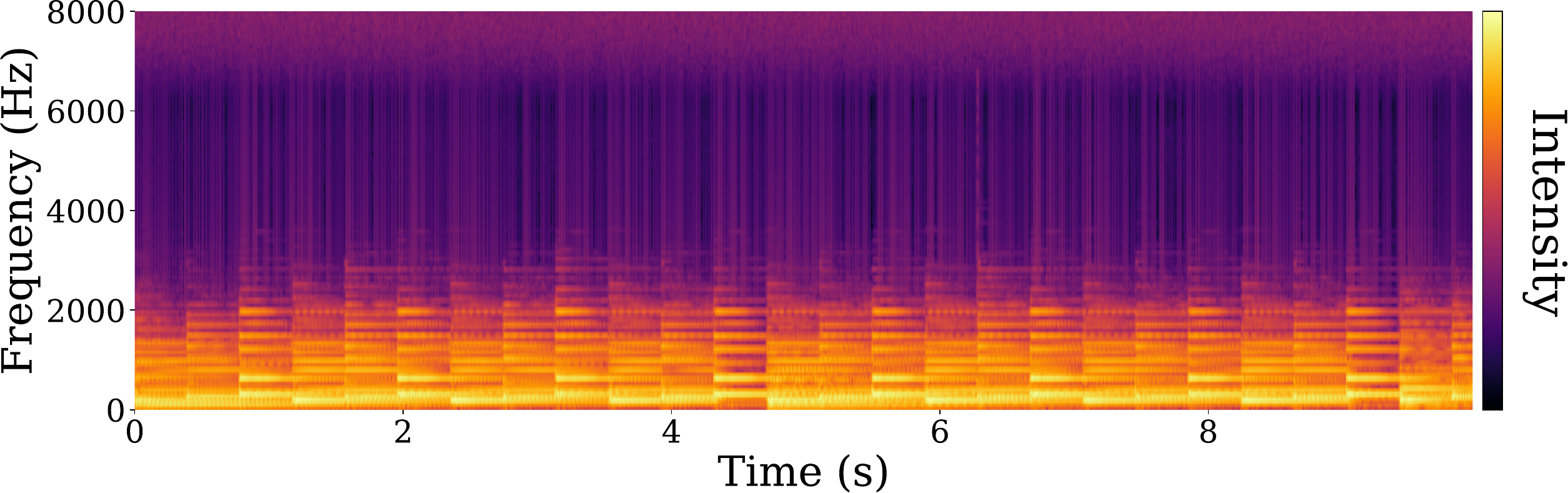}
    \caption{Log--Mel Spectrogram representation of the Bethoveen's piano composition ``Moonlight Sonata''}
    \label{fig:mel_spec}
\end{figure}

A variation of the STFT called Mel spectrogram is commonly used in acoustic analysis and music generation.
Mel spectrogram is based on the STFT spectrogram, but in addition the frequencies are converted into Mel scale. 
The Mel scale aims to simulate the non-linear perception of the human auditory with respect to the sounds, since the energy in a critical band of a frequency has influence in the human hearing. 
The information is transformed into the Mel scale domain in order to extract a set of critical frequency bands by applying a bandpass filter adjusted around the center frequency.
Figure~\ref{fig:mel_spec} shows the log--Mel spectrogram representation of the ``Moonlight Sonata''.

In addition to simulate the perception of the human hearing, some studies consider the use of Mel spectrograms since they can improve the performance by creating more separation between the lower harmonic frequencies~\citep{engel2019gansynth,dhariwal2020jukebox,zhuang2020m2d}.
It allows to avoid the cause blurring and overlap in the lower frequencies.

\section{Chromagram}

This kind of spectrogram yields with the observation that the pitch can be separate into two components: chroma and tone height. 
Chroma is a compressed representation the tonal content in a music piece, where is assumed a tempered scale and independent of the octave according to the C--major scale.
It is represented by a set of twelve pitch spelling attributes (see Table~\ref{tab:pitch_classes}) as the ones used in Western music notation~\citep{christensen2006chwm}. 

\begin{table}[!htpb]
\caption{Chroma descriptors with theirs corresponfing Solfège}
\resizebox{\linewidth}{!}{
\begin{tabular}{lcccccccccccc}
\hline
\hline
\multicolumn{1}{c}{\textbf{\begin{tabular}[c]{@{}c@{}}Chroma \\ \phantom{1}\end{tabular}}} & \begin{tabular}[c]{@{}c@{}}C\\ \phantom{1}\end{tabular} & \begin{tabular}[c]{@{}c@{}}C\# \\ \phantom{1}\end{tabular}     & \begin{tabular}[c]{@{}c@{}}D\\ \phantom{1}\end{tabular}  & \begin{tabular}[c]{@{}c@{}}D\#\\ \phantom{1}\end{tabular}      & \begin{tabular}[c]{@{}c@{}}E \\ \phantom{1}\end{tabular} & \begin{tabular}[c]{@{}c@{}}F\\ \phantom{1}\end{tabular}  & \begin{tabular}[c]{@{}c@{}}F\# \\ \phantom{1}\end{tabular}      & \begin{tabular}[c]{@{}c@{}}G\\ \phantom{1}\end{tabular}   & \begin{tabular}[c]{@{}c@{}}G\#\\ \phantom{1}\end{tabular}      & \begin{tabular}[c]{@{}c@{}}A \\ \phantom{1}\end{tabular} & \begin{tabular}[c]{@{}c@{}}A\#\\ \phantom{1}\end{tabular}      & \begin{tabular}[c]{@{}c@{}}B\\ \phantom{1}\end{tabular}  \\
\textbf{Solfège}                                                                    & do & \begin{tabular}[c]{@{}c@{}}re \\dièse\end{tabular}  & re & \begin{tabular}[c]{@{}c@{}}mi \\dièse\end{tabular} & mi & fa & \begin{tabular}[c]{@{}c@{}}fa \\dièse\end{tabular} & sol & \begin{tabular}[c]{@{}c@{}}sol \\dièse\end{tabular} & la & \begin{tabular}[c]{@{}c@{}}la \\dièse\end{tabular} & ti \\ \hline \hline
\end{tabular}}
\label{tab:pitch_classes}
\end{table}

\begin{figure}[!htpb]
    \centering
    \includegraphics[width=\linewidth]{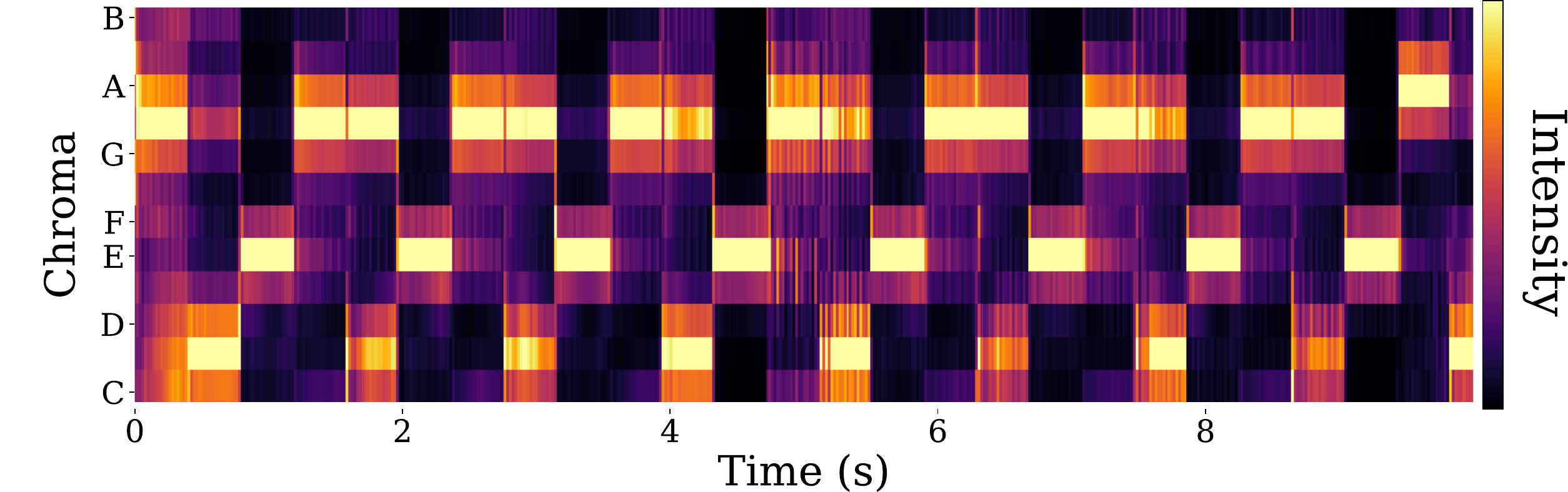}
    \caption{Chromagram representation of the Bethoveen's piano composition ``Moonlight Sonata''}
    \label{fig:chroma_spec}
\end{figure}

The chromagram is a 2-dimensional visual representation of the resulting time--chroma.
These sequence of chroma descriptor represents how the pitch content is spread over the twelve chroma bands within the time window.
An example of the chromagram repsentation of the ``Moonlight Sonata'' is shown in Figure~\ref{fig:chroma_spec}.
The y--axis represents the twelve chromas or pitch classes and the x--axis represent the time in seconds. 

Chroma representation allows to easily interpret music pieces in the frequency domain in which different pitches can be meaningfully categorized, starting from the assumption that humans perceive musical tones as similar in color when they vary by one octave.
Thereby, the chromagram representation has been widely considered in several approaches~\citep{zhuang2020m2d,chen2019ednnsm} in order to improve chord recognition or harmonic similarity estimation.

\section{Data-driven Features, Embeddings}

A representation of music can be transferred from its original form into more advanced features or embeddings.
Word2vec is a technique for natural language processing which is used to train semantic vector spaces based on a dataset of text. 
Each adjacent vector, consisting of hundreds of dimensions, represents a word with a similar context.
Even though music and text are fundamentally different, researchers have applied the concept of word2vec in music to learn vector embeddings.
A skip-gram model was employed by \cite{chuan2020context} to create a semantic vector space for polyphonic music.
Based on the cosine distance, chord relationships and harmonic assignments can be distinguished in music without providing additional input for this purpose.
Further application was implemented in chord2vec \citep{madjiheurem2016chord2vec}, learning representations as vectors of chords.

\chapter{Statistical Modeling}
If the AI is to imitate music styles, the deep learning approach is chosen. 
With statistical modeling, however, training is done with other learning approaches, which require fewer learning processes and develop special degrees of freedom.
Markov chains and Hidden Markov Models are statistical models that can be trained much faster, easier and less data intensive.
Probabilistic context-free grammars further expand the modelling capabilities by recursive structures.

\section{Markov Chains}
A Markov chain (MC) is a statistical model making statements about the probabilities of sequences of random variables. 
These random variables are called states, each of which can take values from a defined set. 
Sets can be words, or tags, as in the speech recognition application, or symbols, which can represent anything, such as the weather or, in case of music, notes.
\\Considering a sequence of state variables $q_{1}, q_{2}, ..., q_{i}$, a Markov model represents the Markov Assumption (1) about the probabilities of this sequence: that the probability of getting into an arbitrary state only depends upon the current state, but not on the previous states. 
\begin{equation} 
P(q_{i} = a|q_{1}...q_{i-1}) = P(q_{i} = a|q_{i-1})
\end{equation}
A stochastic process which fulfils the Markov Assumption is called a first order Markov chain.
An $n$th-order Markov chain is a stochastic process that satisfies the following condition:
\begin{equation} 
P(q_{i} = a|q_{1}...q_{i-1}) = P(q_{i} = a|q_{i-1}...q_{i-n})
\end{equation}
The probability of reaching the next state in this process depends on the $n$ previous states. 
Transitions between states are annotated with probabilities, which indicate the chance that a certain state change might occur.

\subsection{Sampling}
In an early publication \citep{Mar97} the musical aspects of melody (pitch-MC), rhythm (note-rest-MC, note-duration-MC), and dynamics (velocity-MC), are each modeled in four separate MCs, which are at the top level of the fitting and simulation process.
This means that there is no external factor influencing the behavior of these chains and, moreover, there is no interaction between them.
These sub-models act independently of each other, so that an assignment of melody, rhythm and dynamics is more or less random.

For each of the sub-models, the simulation starts with a random initial state. 
At each iteration, the entry of the current state in the transition probability matrix is then used to determine the next state.
To allow external events to influence the improvisation, Marom used a ``Controlled Markov Model'', where both hidden states and observations are observable. 
Sampling is then performed for each underlying state that occurs in the external events, to ``drive'' the improvisation.
Although statistically the music was affected by the control signal, this degree was only slight in the results.
To create solos where the pitch and rhythm are not independent, \citep{Hid17} presents a pattern-based algorithm developed by \citep{NMS13}. 
This algorithm employs Markov chains to randomly generate both melodic and rhythmic patterns using probabilities extracted from the analysis of the corpus of jazz saxophonist.
This means that they have interval vectors for both melodic and rhythmic aspects, which can also be linked together, since each pitch vector is associated with a set of rhythm vectors and vice versa.
To generate a sequence, the algorithm starts by arbitrarily choosing the initial state and then chooses between one of the adjacent states equally arbitrarily.

\section{Hidden Markov Models}
The extension of the Markov chain is the Hidden Markov Model, which is usually defined as a stochastic finite state machine. 
Unlike regular Markov chains, the states $Q$ are hidden, i.e., they cannot be observed directly. 
As with Markov chains, the transition probabilities $A$ indicate the probability of state transitions occurring.
These probabilities, as well as the initial state probabilities $\pi$, are discrete. 
Each state has a set of possible emissions $V$ and a discrete or continuous probability distribution $B$ for those emissions. 
Unlike states, emissions can be observed and thus provide some information about the hidden states, such as the most likely underlying hidden state sequence that led to a particular observation $O$. 
This is called the decoding problem which, together with the evaluation problem and the learning problem, is one of the three main problems formulated for HMMs \citep{Rab86}. There are three algorithms for solving the fundamental problems of HMMs: the Viterbi algorithm (decoding), the forward algorithm (evaluation), and the forward-backward algorithm (learning).
A first-order HMM fulfills the first-order Markov assumption.
Moreover, it instantiates a further assumption that the probability of an initial observation $o_{i}$ depends only on the state that generated the observation $q_{i}$, and not on other states or other observations (Output Independence):
\begin{equation} 
P(o_{i}|q_{1}...q_{i},...,q_{T} ,o_{1},...,o_{i},...,o_{T}) = P(o_{i}|q_{i})
\end{equation}

\subsection{Sampling}
\begin{figure}[ht]
\includegraphics[width=.8\linewidth]{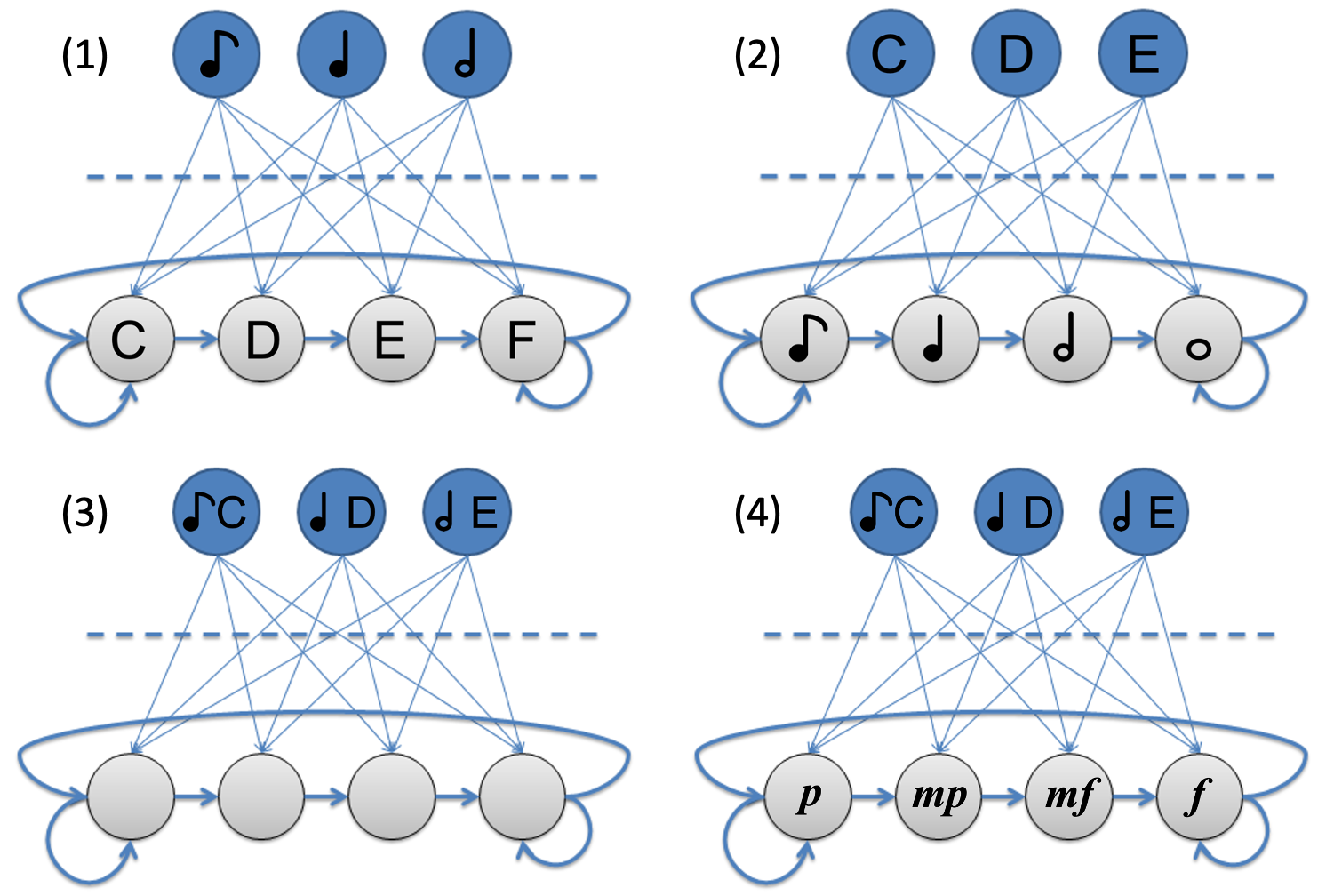}
\caption[Simplified illustrations of HMM layouts]{Simplified representation of the proposed HMM layouts: note-duration layout (1), duration-note layout (2), unassigned-joint layout (3) and velocity-joint layout (4). Background HMM from \citep{Sou10}.}
\label{layouts}
\end{figure}

Music modeling using HMMs requires finding ways to appropriately map the musical dimensions, such as melody, rhythm and dynamics, to the model parameters ($Q$, $A$, $O$, $B$, $\pi$).
Representations for the musical elements in the HMM can be MIDI numbers, semitones and note intervals for melody, classical time units (ms, s), beats and time intervals for rhythm, and MIDI velocity and velocity intervals for dynamics.
\citep{Kat15} combines a HMM with Categorical Distribution (CD) to handle both notes and rhythm together.
The HMM is then divided into two main parts: the MM which is responsible for the notes and the output distribution (CD) which is responsible for the rhythm.
For the rhythm generation problem, a separate CD is used for the duration  of all unique pitches and rests present in the reference music.
However, the assignment of musical elements to the HMM parameters $Q$ and $O$ need not be fixed in this way.
In what follows, we refer to the way musical elements are assigned to states and observations as layout.
An overview of some proposed layouts can be found in Fig. \ref{layouts} as simplified illustrations.
The assignment from \citep{Kat15} is taken as the first layout and from this a second layout results by its inversion, i.e. the rhythmic symbols are in the states and the melodic symbols in the observations.
Third, there is the joint layout that offers the possibility to combine several musical elements in one symbol.
This way, models can be created that have joint notes and durations in the observations and a set of states without any assignment.
The unassigned states of the process could then correspond to an abstract state of mind, such as mood and feeling.
To give the states in the joint layout an assignment as well, velocity is used in the fourth layout.
The velocity-joint layout keeps the joint observations as before and sets velocity to the states, thus bringing melody, rhythm, and dynamics together in one model.
By inversion and further joint combinations of the musical elements, further layouts can be created.
\\After modeling the HMM parameters $Q$ and $O$, follows the initialization of the remaining parameters $A$, $B$ and $\pi$. The initial probability conditions are very important for the learning algorithm and form the starting point for it.
Assuming that the number of states $N$ and the number of observations $M$ are given, we obtain an $N$-dimensional list for the initial state probabilities, an $N\times N$-dimensional matrix for the transition probabilities, and an $N\times M$-dimensional matrix for the emission probabilities to be initialized. 

One way of initialization is presented by \citep{SMB08}, where these probabilities are pre-trained from a data corpus of lead sheets.
To create accompanied (homophonic) music they use a HMM essentially representing which notes frequently co-occur with each chord type (melody observation matrix), and which chords typically precede and follow other chords (chord transition matrix) in the database. 
For this purpose, all probabilities are calculated using maximum likelihood estimation (MLE) followed by softmax-normalization.
The likelihood that the observed distribution of notes occurred under the assumption that a chord is played, is computed by forming the dot product of the observation vector with the logarithm of the corresponding row of the melody observation matrix. This gives the log-likelihood for that chord.
\\Since it is very unlikely that certain notes will appear with certain chords, some combinations of notes and chords will have no observed data. The same applies to the model of \citep{Kat15} which clearly shows that the presence of a symbol in the generated sequence may or may not be present during further generation, however, there is no chance to get a symbol that was not present in the reference melody.
For this reason, a conjugate prior is introduced in \citep{SMB08}.
This adds ``imaginary'' instances of each state for each observation, and we propose to also add ``imaginary'' transitions between each state. 
These imaginary values are very small relative to all observed values in the database. 
This has the effect of removing zeros in the matrices and smoothing the distribution somewhat. 
When using ``imaginary'' transitions, it also allows to simulate a quasi non-fully connected HMM with the used architecture, where the transitions actually exist, but are extremely low weighted.
\\Initialization can also be done with a random, Gaussian or discrete uniform distribution for $A$, $B$, and $\pi$.
Pre-training can be performed on top of these initialization types. 
This results in different combinations for the final initialization of $A$, $B$ and $\pi$.
Finally, we introduce a flexible initialization type that is also pre-trained, but the difference from the initialization types presented so far is that there is no fixed size for $N$ and $M$.
All HMM parameters are empty at the beginning of the pre-training and grow with the training data, with $N$ and $M$ increasing. 
This allows for even more adaptation to the input.
\\After the essential components of music have been mapped to a model, there are several ways to generate music using that model.
For Markov models, two functions are relevant for the direct generation of music: \textit{predict} and \textit{sample}. 
While \textit{predict} aims to predict a sequence of states based on observations, this approach uses \textit{sample}, which draws a more or less random sequence of states with corresponding observations from the model.
For live adaptation and generation of music continuous learning and continuous sampling play an important role, which are achieved by regularly fitting the model and regularly sampling from the model, respectively.
From a musical point of view, it makes sense to use musical events or musical time units, instead of classical time units, to trigger retraining.
In this sense, time-based triggering can be based on the musical time unit of a beat and event-based triggering can be based on the musical event of a note.
These triggers initiate the retraining process, which involves updating the model parameters $A$ and $B$.
In standard HMM training, the transition probabilities and emission probabilities are determined indirectly by the observations using EM algorithm.
This is called unsupervised learning for HMMs, which is useful when the states are hidden and cannot be observed.
In some of the proposed modeling cases, however, both the observations and the states are given, since both notes and note durations can be observed in the player's input sequence.
This allows supervised learning for HMMs, where the model parameters are trained directly using the given state alignment and maximum likelihood estimation as proposed for pre-training.
By continuously invoking one of the two training methods, the HMM adapts live to the current input.
\\How much it adapts can be determined by a weighting parameter.
For this purpose, smoothed weighting is applied as proposed in \citep{Schuk95}, where the a posteriori estimates are obtained by interpolating the ML (machine learning) parameters with the preset or pre-learned model parameters.
With continuous learning, in each retraining, the a posteriori estimated values of the previous time step become the a priori estimated values of the current time step.
In simple terms, to obtain the probability matrix $M_{t}$ for $A$ or $B$ at the current time step $t$, the matrix $M_{t-1}$ from the previous step (initialization or retraining) is interpolated with the matrix $M_{fit}$ from the current training step using the weighting parameter $w$ as follows:
\begin{equation}
\label{weight}
M_{t} = (1 - w) \cdot M_{t-1} + w \cdot M_{fit} \quad 0 \leq w \leq 1
\end{equation}
Thus, in terms of the a posteriori expectation, the more lush the training material on the statistic $M_{fit}$ is seeded, the closer $M_{t}$ nestles to the ML estimates. 
This means that for $w$ equal to 0 no retraining takes place at all and only the a priori probabilities are adopted. 
As $w$ increases, the new matrix $M_{t}$ approximates the current training material more and more, with equal weighting of $M_{t-1}$ and $M_{fit}$ occurring at 0.5. 
When $w$ equals 1, the statistics of the current training material are weighted 100 percent, completely replacing the previous model parameters. 
This results in the HMM forgetting everything it has learned before and adapting completely to the input. 
A music generation based on this would more or less copy the input.
\\Another important aspect of training is, of course, the training data itself.
Regardless of what is being trained, in continuous learning it matters how much past data is used for retraining. 
Limiting the data considered can be done by a sliding window of certain size that is placed over the data.
This gives an additional control parameter to influence the music generation that correlates with the other parameters.
With a small window size and a high weighting parameter, for example, the HMM can be controlled to adapt more to the current theme, while with a very large window size and a low weighting parameter, it can be controlled to learn more of the overall impression of the piece of music.
\\We refer to both musical and statistical sampling, since we take samples from the HMM and use them to synthesize music.
In general, HMM sampling involves drawing a random state sequence and the corresponding feature matrix from the model.
In this way, i.e. pairs of notes with their associated note lengths can be obtained, concatenated and played back.
The number of samples can be specified to generate a piece of music of a particular length.
In this way, with a large number of samples, a whole piece of music can be created, however, for live music generation things are more difficult. 
In addition to continuous learning, continuous sampling must be performed by repeatedly drawing samples from the HMM to interact with the player.
As with training, there can be time-based and event-based triggers for sampling.
The sampling rate, as well as the number of samples generated at each time point, in addition to the parameters of continuous learning, affect the final musical result.

\section{Formal Grammars}
A formal grammar is a linguistic model developed in the field of formal language theory in the 1950s.
In the 1960s they saw their first applications to music, mostly to aid in automated analysis of the structure of classical pieces.
A formal grammar provides a rule system that describes how syntactically valid sentences in a certain (formal) language may be constructed (or analysed) by systematic combination of symbolic words in its vocabulary.

\subsection{Definition}
A formal (context-free) grammar consists at least of four things:
\begin{enumerate}
    \item A set of \emph{terminal symbols} $T$.
    \item A set of \emph{non-terminal symbols} $N$.
    \item A set of \emph{production rules} $P$.
    \item A designated \emph{start symbol} $S$ of the non-terminals.
\end{enumerate}
The terminal symbols are the vocabulary of the language being modeled and thus are the building blocks for valid sentences.
The non-terminal symbols resemble abstract, intermediate steps in the production of a valid sentence.
The production rules describe how non-terminal symbols may be expanded into sequences of (other) (non-)terminal symbols to eventually produce valid sentences.
The start symbol designates a special non-terminal symbol from which the production process must start.
\citep{roads1979grm} deliver a concise yet comprehensive overview of grammar theory fundamentals.

\subsection{Example}
\begin{figure}[ht]
    \centering
    \includegraphics[width=.7\linewidth]{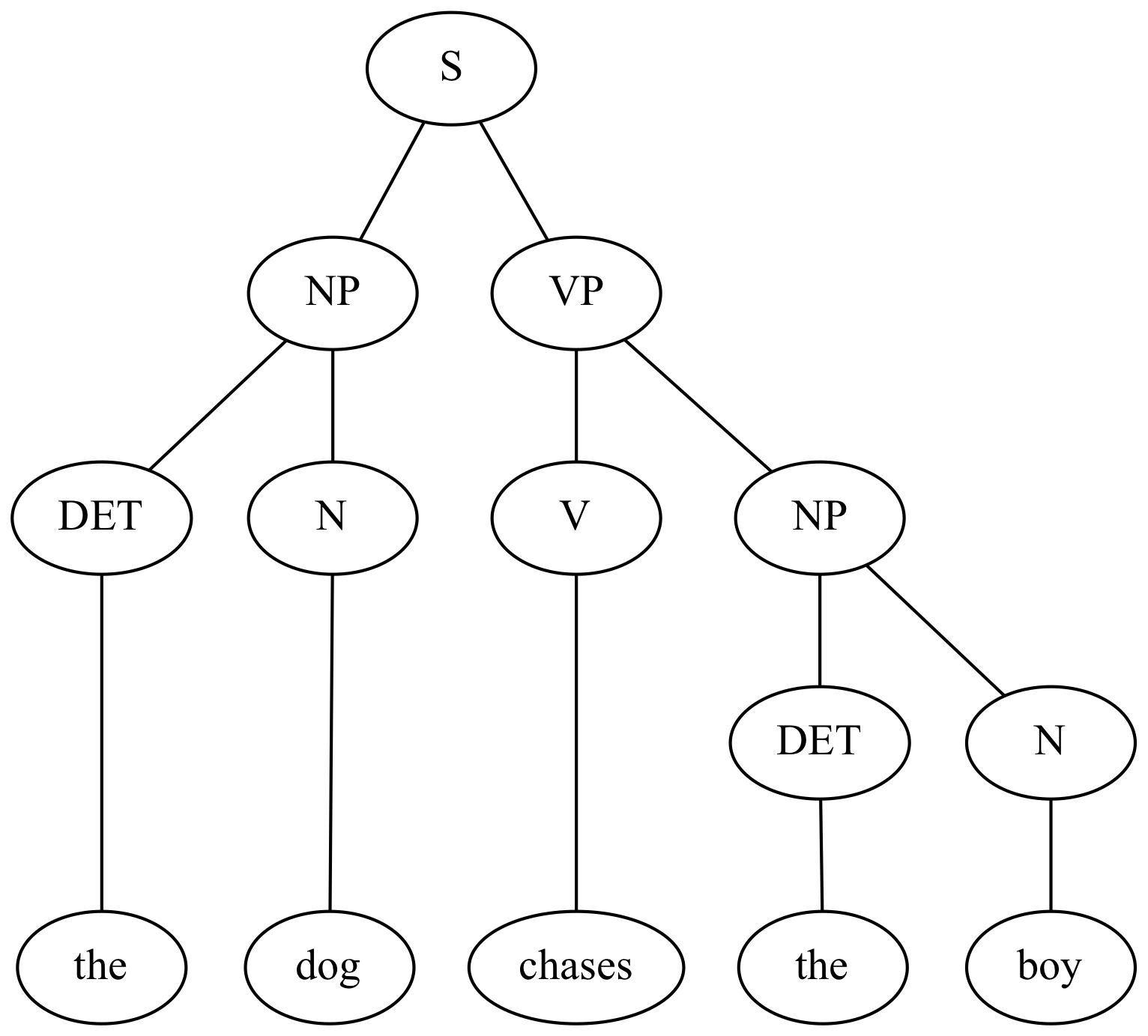}
    \caption{A construction tree for the sentence \enquote{the dog chases the boy}.}
    \label{fig:example_parse_tree}
\end{figure}

The terminal symbols, comma-separated:
\begin{center}
    the , boy , dog , chases 
\end{center}
The non-terminal symbols, comma-separated:
\begin{center}
    S , NP , DET , N , VP , V
\end{center}
The production rules:
\begin{enumerate}
    \item S $\rightarrow$ NP VP
    \item NP $\rightarrow$ DET N
    \item DET $\rightarrow$ the
    \item N $\rightarrow$ boy
    \item N $\rightarrow$ dog
    \item VP $\rightarrow$ V NP
    \item V $\rightarrow$ chases
\end{enumerate}
The start symbol:
\begin{center}
    S
\end{center}
The production rules state that...
\begin{enumerate}
    \item ...a sentence comprises a noun phrase followed by a verb phrase.
    \item ...a noun phrase comprises a determiner followed by a noun.
    \item ...the only valid determiner here is \enquote{the}.
    \item ...a noun may either be expanded to \enquote{boy}...
    \item ...or to \enquote{dog}.
    \item ...a verb phrase comprises a verb followed by a noun phrase.
    \item ...the only valid verb here is \enquote{chases}.
\end{enumerate}
With this simple example grammar, only two valid sentences can be constructed, namely \enquote{the dog chases the boy} and \enquote{the boy chases the dog}.
The systematic application of production rules can readily be visualized as a tree-like diagram, called \emph{construction tree} or \emph{parse tree}, with the start symbol as root, the non-terminals as inner nodes and the terminals as leafs, as shown in figure \ref{fig:example_parse_tree}.

\subsection{Grammars as Abstract Mechanisms}
The presented notion of a grammar as a tool to model valid sentences of a language can be further generalized.
In fact, a sentence is nothing but a sequence of words.
Thus, formal grammars may model any sequence of distinct symbolic elements when carefully choosing the sets of symbols and productions.
Monophonic music may, for example, be modeled as sequence of tones and rests structured as bars and more elaborate abstract groupings of bars that describe the parts of a piece.

Furthermore, formal grammars may be used in two ways.
They may either serve as a tool for construction (a.k.a. generation) to systematically build up sequences that embody a certain logical structure.
Or they may serve as a tool to analyse and explicate (a.k.a. parsing) this logical structure inherent in valid sequences.

\subsection{Probabilistic Grammars}
A probabilistic (context-free) grammar (PCFG) is a formal grammar that resolves ambiguities in its production rules by probabilistic distribution.
Each production rule is assigned a probability such that all production rules' probabilities with the same left-hand side sum to 1, thus forming a distribution.
If, during the production process, an ambiguous production rule is encountered \ie there are additional rules with the same left-hand side, one of them is chosen at random according to their assigned probabilities.

The probabilities themselves may be learned in a supervised fashion.
Given a certain formal grammar, from a corpus of utterances, the most fitting parse tree for each utterance is constructed against that grammar by having a human intuitively resolve its ambiguities.
From here on, the occurrences of all ambiguous production rules are counted against their respective total to form their learned probabilities with respect to the supervisor and the corpus chosen.

\citep{geman2002pgta} give a more thorough introduction to PCFGs and point out the equivalence of various probabilistic grammar types and other statistical models like branching processes and the aforementioned HMMs.

\subsection{Modeling Music with Formal Grammars}
When modeling music with formal grammars, one usually encodes some notion of music theory (\eg tonal harmony) as a formal grammar and probabilistically constructs musical utterances from them.
Since no universal structure of music has (at least yet) been discovered, musical grammars tend to deliver compelling results in rather limited stylistic domains with somewhat strict notions of rhythmic and harmonic structure \eg baroque or swing.

The most generally applicable approach in this sense has been made by \citep{lerdahl1983gttm} in their \emph{Generative Theory of Tonal Music} (GTTM).
In their landmark work, they strive -- in a true Chomskian sense -- a "formal description of the musical intuitions of a listener who is experienced in a musical idiom" thus stating a true, natural way of how music is implicitly structured for a listener with western cultural upbringing.

More concrete and quite early approaches of modeling music with formal grammars are presented by \citep{roads1979grm}.
\citep{keller2007gaai} developed a software tool for automatic generation of convincing jazz solos by modeling characteristic tone relations and rhythmic figures as a probabilistic grammar. They also deliver a quite comprehensive literature overview regarding previous efforts.
\citep{gillick2010mljg} further extend the latter approach by machine-learning the jazz grammars.

\chapter{Parametric Modeling}\label{sec:parametric_modeling}


Any model that captures all the information about its predictions within a finite set of parameters. Given the multiple definitions of the word model a parametric model can output either a probability or a value (in some cases a classification). While non-parametric models assume that the data  distribution cannot be defined in terms of such a finite set of parameters. But they can often be  defined  by assuming an infinite dimensional. 
Usually we think of as a function.
Most of these approaches lies into the concept of machine learning, where deep learning models have be renowned by their ability of learning complex patterns.
Nowadays, automatic music generation encompass several types of Deep Neural Networks (DNN) that goes from analyzing just the time series to trying to model context and melody at the same time. 
\section{Recurrent Models for Music Accompaniment}

Most related works address the problem of algorithmic music compositions via RNNs.
This kind of network aims to process sequential data in deep neural networks, which allows the generation of music sequentially.
The decision in a time instant $t-1$ affects the decision of an RNN in a time instant $t$. 
The networks are connected to their past decisions (feedback loop), where the sequential information is stored in the hidden state $h$ of a network, which manages to span many time steps.
The incorporation of the feed-back allows that the network to have ``memory''.
There are three typical variations of the RNN-based architectures which are vanilla-RNNs, Long Short-Term Memory (LSTM) and Gated Recurrent Units (GRU). 
LSTMs and GRUs architectures consist of internal mechanisms called gates, which attempts to regulate the flow of information, and were created as the solution to short-term memory in vanilla-RNNs. 
These gates aims to learn what data in a given sequence is important to keep or discard.
In~\citep{chung2014egrnsm}, some experiments for music generations using these three diferent RNN architectures performed closely to each other, however, the evaluation demonstrated the superiority of the gated units over vanilla-RNNs.

In the past few years, these Sequence to Sequence (Seq2Seq) approaches have been used to predict the sequence of notes that should accompany a given input.
RNN architectures are widely considered in several approaches for music accompaniment, in which the melody example is used to generate an accompaniment.
One of the challenging tasks nowadays is harmonization, which is in charge to provide accompaniment to an existing melody. 
This task was addressed several studies~\citep{lyu2015mhdslstm,liang2017ascbc,raja2020mgtsa}, where different sequential encoding schemes based on LSTM architectures were proposed to model composition and harmonization, however these approaches were limited to pitch and rhythm, avoiding encoding music theoretic concepts.
For instance, in~\citep{de2019rcm}, the authors proposed a methodology, which consists on generating the melody according to predefined rhythmic and harmony templates. 

Symbolic generation is currently the most prevalent approach in which RNN based models are used.
These models train and generate at the note level, and combining with LSTMs cells appear to model common music theoretic concepts without prior knowledge~\citep{liang2017ascbc}.
For instance, some studies such as ``BachBot''~\citep{liang2017ascbc} and ``MelodyRNN''~\citep{waite2016gltss}, claim that they can generate realistic melodies inspired in Bach Chorales as same as model some aspects related to harmonization by using LSTMs.
``BachBot'' was development as an end-to-end automatic composition system based on a LSTM approach for accompaniment and composing music.
The ``MelodyRNN'' model was proposed by the Google Magenta project to generate a melody sequence from a priming melody.
In~\citep{ren2020popmag} the authors demonstrated that RNNs can generate multiple sequences simultaneously considering hierarchical RNN models to generate the melody, the drums, and chords, leading to a multitrack pop song.

Follow-up research has been made specially to generate accompaniments from symbolic sources using classical LSTMs~\citep{chung2014egrnsm,kalingeri2016mgdl,liang2017ascbc,agarwal2018lmgrt}, Biaxial LSTMs~\citep{johnson2017gpmtpn,mao2018deepj,mangal2019lbmgs}, Tied Parallel LSTM with Neural Autoregressive Distribution Estimator (TP-LSTM-NADE)~\citep{johnson2017gpmtpn}, GRUs~\citep{chung2014egrnsm} and combined RNN based architectures with other different DNN models~\citep{lyu2015mhdslstm,raja2020mgtsa}.
These models capture long-term dependencies of melodic structure, however, have difficulty capturing and learning some intrinsic components of a raw audio signal. 

Some related works~\citep{kalingeri2016mgdl,defossez2018sing} are focused on unconstrained music generation, which means that there is not  information about musical structure directly involved to aid learning.
Raw signal or spectrograms representations are also considered to capture features in the frequency domain and increase the quality of music generated.
Usually, the RNNs using raw signals are combined with other type of DNN architectures based on Restricted Boltzman Machines (RBM), Convolutional Neural Network (CNN)~\citep{kalingeri2016mgdl,manzelli2018cdgam,manzelli2018eteamg,defossez2018sing}.
This allows the production of realistic sounds, although unstructured music.
Existing automatic music generation approaches~\citep{manzelli2018cdgam,manzelli2018eteamg} attempt to combine the benefits of extract information form symbolic and audio representations.
Commonly, RNNs are applied to learn the melodic structure of different styles of music from a symbolic output and other different architectures such as WaveNet or CNNs, which incorporate some additional aspects of the composition as a secondary input to the network.

\subsection{Inference and control}
In the case of sampling, recurrent systems differ according to the purpose for which they were designed.
The sequences, whether of the same or varying length, are first embedded in sequences of tokens of an n-dimensional vector space.
Regardless of the underlying recurrent model, the network is propagated and is usually passed through a softmax in the final layer to obtain a predictive distribution $P(x_{t+1}|h_{t-1},x_{t})$. 
This corresponds to the probability distribution for the next token $x_{t+1}$ considering the token $x_t$ and the state of the memory cell state $h_{t-1}$.
The predicted token with the highest probability is the model output and also represents the input for the next iteration.
The auto-regressive approach generates the final sequence token by token.

Different sampling strategies can be applied to the conditional probability. 
By selecting the token with the highest probability it corresponds to a 1-best search at a time step t, e.g. $argmax_{x_{t}}P(x_{t}|x_{<t})$.
This produces a deterministic result for the same input.
In a non-deterministic strategy, the most likely token is not selected, but can be determined, for example, from the n-best based on a certain threshold.

Loopback RNN \citep{waite2016gltss} introduces special inputs and labels to detect patterns that occur across 1 and 2 bars.
In addition to the previous event, additional events from 1 and 2 bars before are fed in with the supplementary information whether the last event is a repetition of the last two bars. 
This allows the model to learn repeating patterns.

In DeepJ \citep{mao2018deepj}, the model's probability distribution is used to sample whether a note is played or not using a coin flip. If a note is played, it is also sampled from the repeat probability whether this note will be played again. No priming takes place, so the complete sequence is generated from scratch. The model allows control over the style of the sample being generated.

In autoregressive systems, the music is always generated from left to right and allows little interaction by the user. \cite{hadjeres2017deepbach} has tried to solve this problem with the LSTM-based model DeepBach. It does not sample and model each part from left to right but contains four neural networks each, two recurrent networks for summing past and future information, one for notes occuring at the same time and a final neural network for merging the outputs from the three previous ones.
DeepBach allows interaction and constraint by the user by imposing positional constraints like notes, rhythms or cadences in the generated score.

A hierarchical RNN for generating pop multi-track music was proposed by \cite{chu2016song}, where each layer produces a key aspect of the whole song. 
The melody is generated in the lower layer, followed by the drums and chords in the layer above.
In addition, conditioning of the model to the scale type is possible. 
The generation is performed for one note in each time step, chords are generated every half bar and also the drum patterns. The length of the sequences to be generated is determined randomly in advance.

\cite{lim2017chord} has proposed a system with two BLSTM network layers to learn the correspondence between melody and chord sequence pairings, with the limitation that one chord can be generated per measure. The evaluations were performed on 4 bars of generated data.

\section{Modeling Context}  

There are other DNN-based approaches that are used more to transform the input data into a new melody given a predefined model.
These approaches frequently use CNNs in different applications in music generation to process graphical representations~\citep{kalingeri2016mgdl,manzelli2018cdgam,manzelli2018eteamg,defossez2018sing,engel2017nas,engel2019gansynth}. 
CNNs are designed to process multiple arrays. These networks are formed by a structure of alternating convolutional filters and pooling layers instead of the fully connected layers of a DNN.
For instance, the analysis of two-dimensional arrays from a time-frequency representation (spectrograms, chromagrams) from audio signals. 
It allows to extract information from those representations to generate a music piece.

Usually, CNN models are integrated to different deep learning approaches in order to generate sequences, probabilistic distributions or other general representations, which aids to obtain more realistic sounds by obtaining information from the raw audio signals.
Convolutional networks along with recurrent and fully connected layers have been widely using to capture information in the frequency domain focusing on unconstrained learning, in which simultaneously, these models aim of capturing temporal dependencies to generate coherent music pieces~\citep{kalingeri2016mgdl,manzelli2018cdgam,manzelli2018eteamg,defossez2018sing}.
The different applications with CNN goes from digital vocoders generating a singer's voice~\citep{bonada2007ssvs} to synthesizers producing timbres for various musical instruments~\citep{engel2017nas,engel2019gansynth}.

Other models that uses CNNs such as WaveNet~\citep{oord2016wavenet}, are currently very active in the state-of-the-art of music generation~\cite{manzelli2018cdgam,manzelli2018eteamg,michelashvili2020htpag}.
This architectures proposed by DeepMind in 2016, is one of the most used networks to generate music from raw signals nowadays. 
It is characterized by being a powerful generative approach to probabilistic modeling of raw audio, which consists of a CNN-based model proposed for creating raw waveforms of speech and music, initially thought for text-to-speech approaches. 
This model uses a series of dilated convolutions which allows the model to exponentially increase the context length.
Several studies applies and are based on this network because it demonstrated that can also generate realistic musical waveforms and have proven to be effective at modeling short and
medium scale signals. 
Biaxial LSTM models are usually combined with Wavenet networks to represent the long-range melodic structure and interpret the generated melodic structure in raw audio form~\citep{manzelli2018cdgam,manzelli2018eteamg}.  
In most of these approaches the notes of the composition acts as a secondary input to the network.
It allows skipping the intermediate interpretation step of the symbolic representations, which also helps to provide a structure from the raw audio to the output, keeping some properties of both models.
Hierarchical architectures based on Wavenet are proposed to learn in different scales patterns related to the pitch and loudness, such that the synthesized audio mimics the timbre and articulation of a target instrument~\citep{michelashvili2020htpag}. 
These kind of models aids to improve pitch coherence and create realistically sounding samples.
WaveNet are also use for addressing the problem of fully automatic music and choreography, such as in Music2Dance~\citep{zhuang2020m2d}, where the authors proposed a methodology to shift the WaveNet for human motion synthesis by considering the characteristics of rhythms and melody.
Synthesis of music is commonly addressed using WaveNet. 
For instance, Wave2Midi2Wave~\citep{hawthorne2018efpfmd} generate synthetic compositions conditioned on a piano roll representation.

\subsection{Inference and control}
Similar to the recurrent models, in CNN systems a softmax activation often is applied in the last layer of the network, by which the probability distribution is represented.
In the case of WaveNet, content is generated sequentially; after each prediction of a sample, it is fed back into the network to predict the next sample. 
Dilated convolutions enable the WaveNet strategy to deal with long-term correlations.
Filters with gaps are used in successive convolutional layers so that a connectivity pattern in tree structure is formed over several layers. 
In this way, dependencies are learned over several time steps, but this is still not sufficient for the ability to capture longer-term musical structures. 
By extending the receptive field to several seconds, musical sounding samples could finally be generated. 
However, the individual parts varied from second to second in genre, instrumentation, volume and sound quality as claimed by the authors.
To overcome this problem, conditional models were trained with conditions on, for example, genre or instruments.
However, the length of the generated examples remains limited to a few seconds and is time-consuming and costly due to the autoregressive approach.

To better represent the long-term structures of music and to avoid the limitations of WaveNet, \cite{dieleman2018challenge} introduced a hierarchical WaveNet. 
The basic idea is to decouple the learning of local and large-scale structures.
The training is split into two separate WaveNet models, for high-level representations and local signal structures conditioned on the high-level representation.
Most of the samples are 10 seconds long, but also minute long samples were generated with the architecture.

Since CNN is used to directly synthesize the audio, systems based on it can usually only reproduce long-term structures of the music to a very limited extent.
In addition, real-time generation is not possible due to the amount of time steps required for audio data.
Control is introduced in CNN systems via conditions, which is mainly for regularization in the long-term structure. 
However, this allows a certain degree of control of the generated examples.
It is mostly implemented with binary vector representations of tags for a training sample. 
Tags can be for example a key, genre, instruments or style. 


\section{Generative Models}
%
This kind of models in music generation aims to learn data distribution, all this applied within the paradigm of unsupervised learning.
The main assumption is that by learning how to generate the data, the best features of the data can be also learned.
Several kinds of deep generative models have been introduced in the past few years. 
In the music generation domain the use of generative models have been seen rapid progress due to algorithmic improvements and the availability of high-quality musical data.
Typically, the symbolic based approaches makes the modeling problem easier by working on the problem in the lower-dimensional space.
Some of these models generated symbolic music, for instance, in the form of a pianoroll, which aims to specify the timing, pitch, velocity, and instrument of each note to be played.
Some success has been achieved with models that are able to produce music pieces either in the raw audio~\citep{} and in the spectral domain~\citep{}.
Nowadays, some of the most well-known generative models used to generate music representations are: AEs~\citep{engel2017nas, roberts2018hlvllsm, kim2019nmsftc, dhariwal2020jukebox}, GANs~\citep{yang2017midinet, dong2018musegan, engel2019gansynth}, and NADE~\citep{johnson2017gpmtpn}.

\begin{figure}[ht]
    \centering
    \includegraphics[width=.7\linewidth]{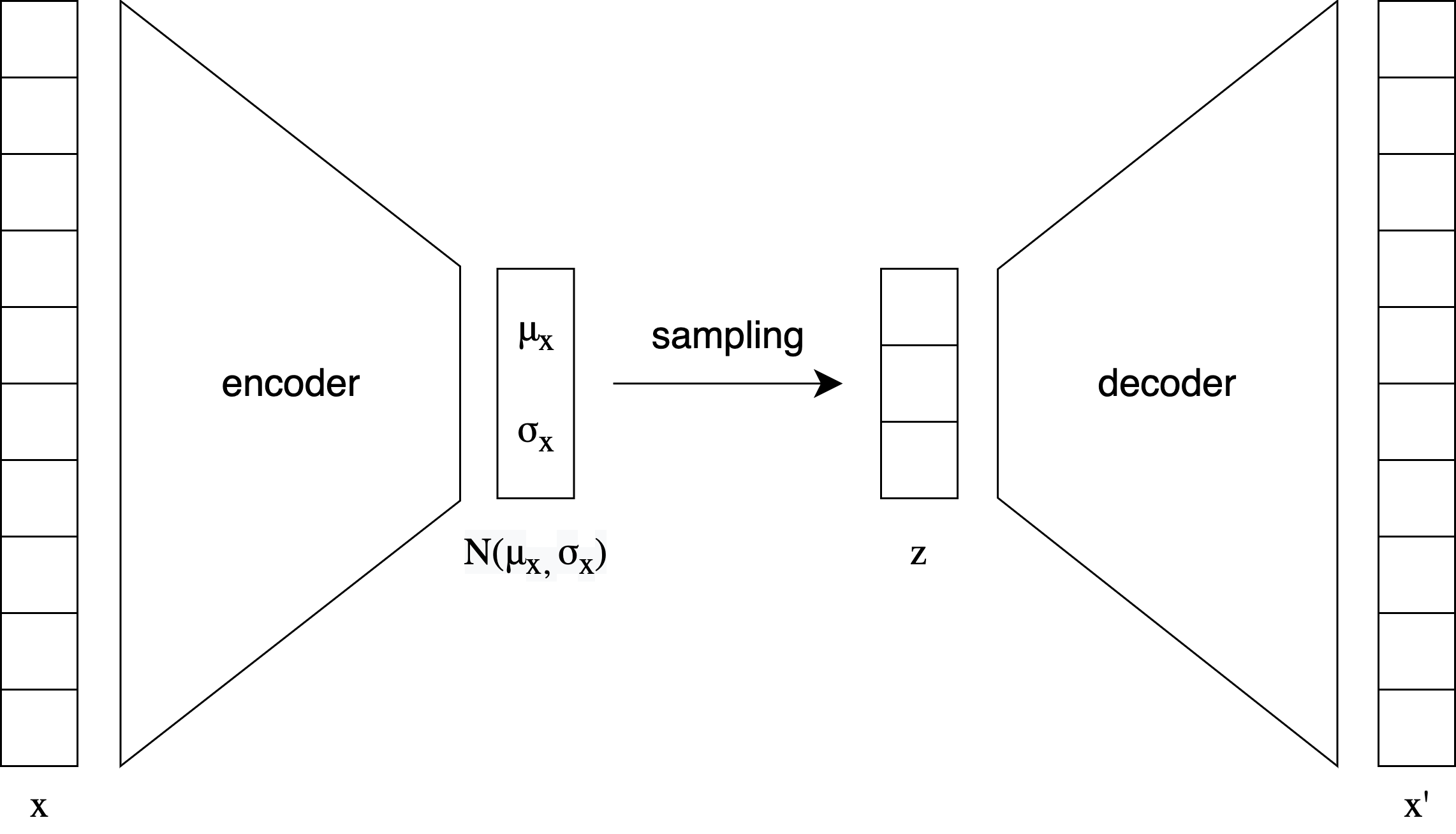}
    \caption{Higher level representation of a VAE consisting of encoder, the Gaussian distributed latent space and decoder.}
    \label{fig:vae}
\end{figure}

AEs are feed-forward networks, which aim to generate new data by compressing the input in a space of latent variables and then reconstructing the output based on the acquired information.
The AE consists of three main parts: (1) encoder, which compresses the information in a space of latent variables called code, and (2) decoder, the part that tries to reconstruct the input based on previously collected information (code).
Variational AE (VAE)~\citep{kingma2014sgvae} provides a Gaussian priors on the latent space, where the encoder is trained to describe a probability distribution for each latent attribute. A higher level representation is shown in Fig. \ref{fig:vae}. Inputs are fed into an encoder network and mapped to a Gaussian distributed latent space $q(z|X)$ parameterized with mean $\mu$ and variance $\sigma$.
The minimization of the error in a VAE consists of two terms in the loss function. The \textit{reconstruction loss}, how well the original input data was reconstructed and the \textit{latent loss}, which is the regularization term. Equation \ref{eq:vaeloss} represents the two terms. The reconstruction loss is measured by using the reconstruction log-likelihood $p(x|z)$. This shows how well the decoder has learned to reconstruct the input $x$ given the latent representation $z$. $D_{KL}$ represents the latent loss which is the Kullback Leibler divergence between the distribution $q(z|x)$ of the encoded input and the normal distribution $p(z)$ with mean 0 and standard distribution of 1.

\begin{equation}
    \label{eq:vaeloss}
    \begin{aligned}
        \mathcal{L} = \mathbb{E}[log p(x|z)] - D_{KL}[q(z|x)||p(z)]
    \end{aligned}
\end{equation}

Some approaches  such as NSynth~\cite{engel2017nas} and Mel2Mel~\citep{kim2019nmsftc} considered a combination of WaveNet and AEs for synthesis of music conditioned on symbolic music information.
AE models for music can also be used for music style transfer. 
In Midi-VAE~\citep{brunner2018midivae}, the authors considered VAEs to transfer styles from classical to jazz music. 
For instance, for the Universal Music Translator Network~\citep{mor2019aebmt} a denoising AE was used to disentangle musical style and content.
VAEs are also used together with hierarchical RNNs to generate music such as in MusicVAE~\citep{roberts2018hlvllsm}.
Mel spectrograms and a multi-scale Vector Quantisation–VAE (VQ-VAE) were combined in Jukebox~\citep{dhariwal2020jukebox} to compress extensive context of raw audio into discrete codes.

\begin{figure}[ht]
    \centering
    \includegraphics[width=0.3\linewidth]{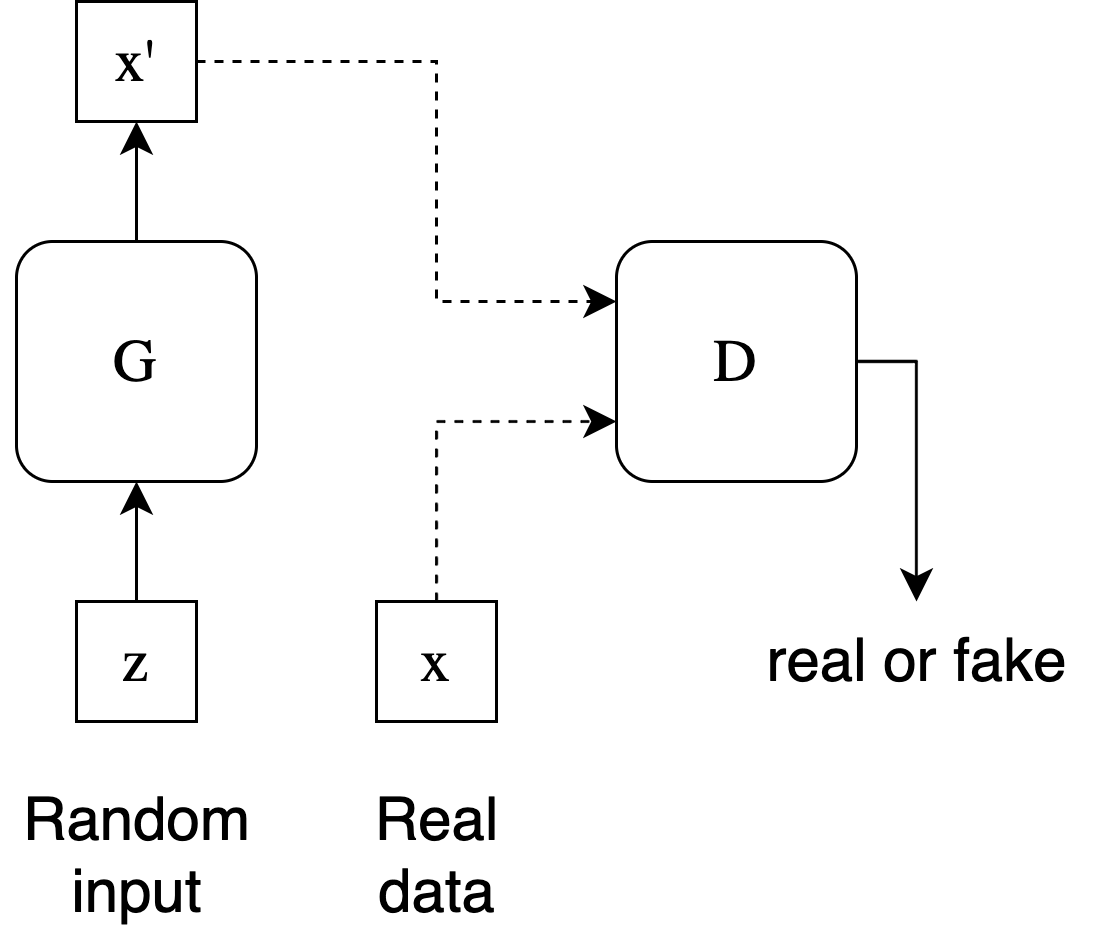}
    \caption{Generative adversarial network (GAN) architecture, reproduced from \cite{huzaifah2020deep}.}
    \label{fig:gan}
\end{figure}

GANs~\citep{goodfellow2014gan} are composed by two feed-forward networks that try to generate samples from "scratch".
These two networks, called generator $G$ and discriminator $D$, compete with each other to make the generated samples as indistinguishable as possible from the data, illustrated in Fig. \ref{fig:gan}.
The goal of the generator $G$ is to transform a random noise vector into a synthetic sample that resembles real samples from a distribution of real data.
The discriminator $D$ plays the role of a classifier, which estimates the probability that a sample comes from real data than from the examples of the generator $G$. The training corresponds to a \textit{minimax} game with a objective where $G$ recovers the distribution of the training data and $D$ always outputs 1/2. Equation \ref{eq:ganobjective} shows the objective of GAN training. $D(x)$ represents the probability that $x$ was drawn from real data. $\mathbb{E}_{x\sim p_{\text{data}}(x)}[\log{D(x)}]$ is the expectation of $log D(x)$ with respect to $x$ being drawn from real data.
$1 - D(G(z))$ represents the probability that $G(z)$ was not drawn from real data. The objective of $D$ is to maximize both expectation terms and thus $V(G,D)$. In opposite, the objective of $G$ is to minimize the value function $V(G,D)$.

They are renowned for their ability to generate high-quality pictures~\cite{}.
GANs are combined to generate single-track or multi-track symbolic music pieces. 
The adversarial idea aids in the generation of long sequences, which enforces the models to focus on learning global and local structures.

\begin{equation}
    \label{eq:ganobjective}
    \begin{aligned}
        \min_{G}\max_{D} V (G, D) = \mathbb{E}_{x\sim p_{\text{data}}(x)}[\log{D(x)}] + \\ \mathbb{E}_{z\sim p_{\text{generated}}(z)}[1 - \log{D(G(z))}]
    \end{aligned}
\end{equation}

The challenge of improving patterns in the melody and chords is addressed via Temporal GANs (TGAN) in~\citep{cheng2020vmtgan}, where the authors additionally proposed a new discriminator to recognize the time sequence of music and use of a pretrained beat generator to improve quality of patterned melodies and chords.
The prediction of rhythm and melody independently using a GAN-based model was addressed in~\citep{liu2018lsgan}, which  the used of two different generator was proposed.
MidiNet~\citep{yang2017midinet} is another GAN based approach that aims to exploit available prior based on melody and chord, where the authors claimed that the proposed model is able to generate melodies from scratch, by following a chord sequence of MIDI notes.
MuseGAN~\citep{dong2018musegan} which use generative adversarial networks three different GAN models (the jamming, the composer model, and the hybrid model) for symbolic multi-track music compositions. 
Automatic generation of synthetic music is addressed in GanSynth~\citep{engel2019gansynth}, which uses GANs to produce magnitude spectrograms together with instantaneous frequencies for easier spectrogram inversion to generate musical notes.
Sequence GANs (SeqGANs) are used for discrete sequence generation and for creating polyphonic musical sequences.
For instance, in~\citep{lee2017pmggan} is proposed a SeqGAN based model to compress the duration, octaves, and keys of melody and chords into a single vector representation.

AEs and GANs are often combined.
MIDI-Sandwich used a multi-model multi-task hierarchical conditional VAE-GAN network to combine musical knowledge.
Later on MIDI-Sandwich 2 was proposed, where one of the modifications of the original model consisted on adding a recurrent component.

\subsection{Inference and control}
In music, generative models are mainly used in the field of generation from scratch.
In the GAN architecture, content is generated after adversarial training of the generator by sampling random latent variables.
Similarly, a VAE can sample a vector from the latent space after the training phase, and the decoder generates iteratively content.
This can be realized by random sampling or by interpolation between values of latent variables belonging to two pieces of music, for example.
Interpolation has been successfully demonstrated in MusicVAE~\citep{roberts2018hlvllsm}.

A GLSR-VAE architecture was proposed by \cite{hadjeres2017glsr} to better control data embedding in latent space.
After determining the geometric structure of the space, the geodesic potential space regularization (GLSR) method is used for regularization. 
The variations in the latent space thus map the variations of data attributes, allowing control of attributes of the generated data.

\cite{tan2020music} proposed a system called Music FaderNets, which first models quantifiable low-level attributes and then learns high-level representations with constrained numbers of data. 
So-called "faders" allow low-level features of the music to be adjusted independently without affecting others.
Finally, Gaussian Mixture Variational Autoencoders (GM-VAEs) are used to infer high-level features from low-level features by semi-supervised clustering.

\cite{teng2017generating} have proposed a hybrid neural network and rule-based system for generating pop music. 
The network consists of a conditional variational recurrent autoencoder.
The overall structure of the music and the chord progressions are generated by a model augmented with a temporal production grammar.
The encoder maps a melody to the multi-dimensional feature space, with conditioning on the underlying chord progression.
A melody is generated by sampling a random vector from the latent space and passing it to the decoder along with a chord progression generated by the grammar.


\section{Attention Based Models}
Attention-based models are used in a wide variety of areas and often achieve state-of-the-art results.
The goal is to mimic cognitive attention and to focus on small but important parts of the data.
A vector of importance weights is formed, whereby for an element, for example a note, an attention vector is used to estimate how strongly it correlates with other elements.

\begin{figure}[ht]
    \centering
    \includegraphics[width=.8\linewidth]{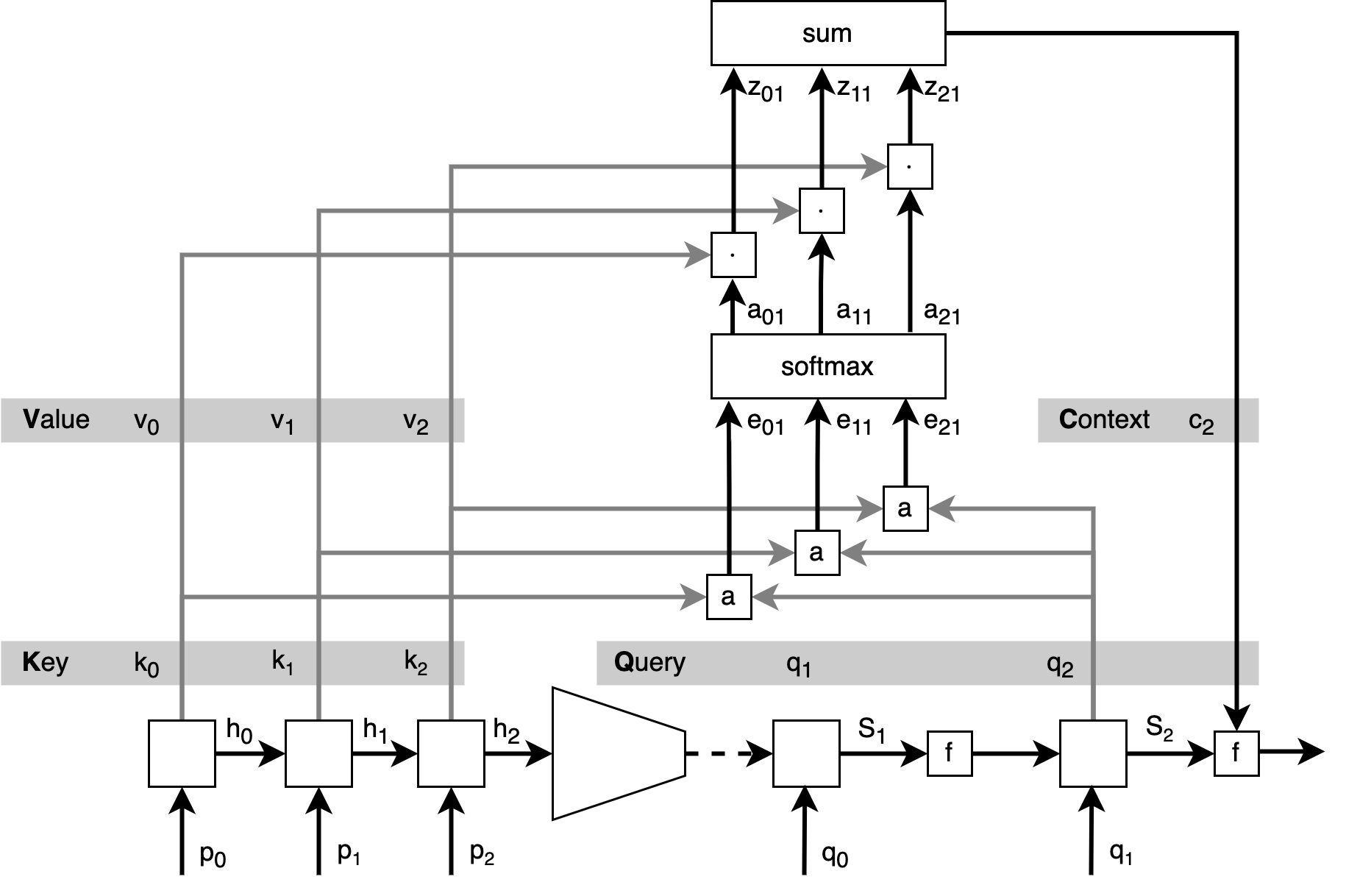}
    \caption{The encoder-decoder attention mechanism visualized.}
    \label{fig:attention}
\end{figure}

The first type of attention introduced by \cite{bahdanau2014neural} is often called Additive Attention, which was aimed at improving a sequence-to-sequence model in machine translation.
The context vector $c_t$from equation \ref{eq:attentioncontextvector} is the sum of the hidden states of the input sequence weighted by the alignment scores.
$\alpha_{ij}$ from equation \ref{eq:attentionweight} shows the amount of attention the $i$th output should put on the $j$th input. This is calculated by taking the softmax over all attention scores, denoted by $e$.
The $e$ often called energy results from the alignment model $a$, which scores how well the inputs around position $j$ and the output at position $i$ align, where $s_{i-1}$ is the hidden state from the previous time step. 
Attention can be generalized by taking, for a given query $Q$ and a set of key-value pairs $(K,V)$, the weighted sum of the values depending on the query and the corresponding keys. 
In the Additive Attention example, the query corresponds to the previous hidden state $s_{i-1}$ and the keys and values to the hidden states.
The attention mechanism is illustrated in Fig. \ref{fig:attention}.

Attention is used in combination with different architectures in combination with for example CNNs and RNNs.

\begin{equation}
    \label{eq:attentioncontextvector}
    \begin{aligned}
        c_i = \sum_{j=1}^{T_x} \alpha_{ij}h_j
    \end{aligned}
\end{equation}

\begin{equation}
    \label{eq:attentionweight}
    \begin{aligned}
        \alpha_{ij} = \frac{\exp(e_{ij})}{\sum_{k=1}^{T_x} \exp(e_{ik})}
    \end{aligned}
\end{equation}

\begin{equation}
    \label{eq:attentionenergy}
    \begin{aligned}
        e_{ij} = a(s_{i-1}, h_j)
    \end{aligned}
\end{equation}

A sequence model based on self-attention is the Transformer \citep{vaswani2017attention}, which consists of a encoder and a decoder.
The main component of the transformer is the multi-head self-attention mechanism.
Instead of computing the attention only once, the multi-head mechanism runs the scaled dot-product attention several times in parallel.
Afterwards, a concatenation and linear transformation is performed to obtain the expected dimension.
The encoder consists of a stack of identical layers, consisting of, among other layers, multi-head self-attention layer followed by a fully connected feed-forward network.
The output of the encoder is an attention-based representation with the information about which parts of the inputs are relevant to each other with a potentially infinite-length context.
The decoder works similar to the encoder, with an additional attention mechanism which draws relevant information from the generated encodings of the encoder.

With the Music Transformer \citep{huang2018music}, an attempt was made to model minute-long music sequences with the help of the transformer. 
Relative attention was employed, which is well suited for generative modeling of symbolic music.
Transformer autoencoder \citep{choi2020encoding} deals with the topic of how to introduce interaction into music generation. 
It extracts MIDI data from .wav files with the Onsets and Frames \citep{hawthorne2018onsets} model and encodes a global representation of the style with the Transformer autoencoder.
In the decoder, the style vector prevents the model from remembering only the reconstruction during training.
MuseNet \citep{payne2019muse} uses the Sparse Transformer \citep{child2019generating} based on GPT-2 to also implement long-term structure in a music piece. 
Due to the full attention over a context of 4096 tokens, the generation of 4-minute pieces is possible. Based on a few notes input, the model can generate music in different styles with several instruments including piano, drums, bass and guitar.
With the Jazz Transformer \citep{wu2020jazz} an attempt was made to generate jazz music based on the Transformer-XL model.
A novel adversarial transformer \citep{zhang2020latfsmg} was proposed to generate musical pieces with high musicality.
It combines adversarial learning with the self-attention architecture to provide a regularization that forces the Transformer to focus on learning global and local structures.

Transformers have also been used to generate multi-track music. The LakhNES \citep{Donahue2019LakhNESIM} is a proposed transformer for multi-instrument music generation that is first pre-trained on one dataset (Lakh MIDI) and then fine-tuned on another (NES-MDB) to improve model performance.
Multi-Track Music Machine (MMM) \citep{ens2020mmm} is based on the Transformer, creating a time-sorted sequence for each track and concatenating multiple tracks into a single sequence.
Pop Music Transformer \citep{huang2020pop}, based on Transformer-XL focuses on music score transformation into an event set, which provides a metric context for modeling the rhythm patterns of the music.


\subsection{Inference and control}
Transformers are trained on sequence data, for a series of notes the subsequent note is predicted.
The data, for example MIDI, must be encoded into suitable tokens for training.
This can be done in different ways, for example a chord can be encoded as a single token or one token for each note of the chord.
For example, in the Music Transformer \citep{huang2018music}, 128 $NOTE\_ON$ and $NOTE\_OFF$ events corresponding to MIDI pitches are used for the Piano-e-Competition dataset. In addition, events for the forward time shift are introduced to represent the expressive timing.
In addition, the time progression must be modeled, this can be for example per token as a musical beat or a fraction of it or a token is marked with the absolute time in seconds.
In the case of the MuseNet \citep{payne2019muse} model, additional embeddings were added to provide more structural context to the model.
A learned embedding was added to the standard positional embeddings, which tracks the passage of time, giving simultaneously sounding notes the same timing embedding. 
In addition, the model allows control in the form of composer/style and the definition of the instruments included.
During training, the composer and instrumentation tokens are prepended to each sample.
The length of the generated sequences is variable and is determined by the number of tokens.

The Sparse Transformer \citep{child2019generating} model and the use of the sparse attention mechanism, which is supposed to scale better to long sequences, was used to generate music in waveforms, among other tasks. 
With sparse attention it was possible to handle waveforms up to 65k time steps, which corresponds to about 5 minutes at 12 kHz.
The Music Transformer \citep{huang2018music} is able to generate long-time structures of 60 seconds (~2000 tokens) for piano performances, however, training is performed on symbolic data. 
Furthermore it is possible with the Music Transformer to generate a complete performance (melody and accompaniment) based on an input melody. 
For this, the melody is used in the form of a quantized sequence of tokens as input to the encoder and the decoder realizes the performance.

Using the Transformer Autoencoder \citep{choi2020encoding}, various conditioning tasks were addressed.
In performance conditioning, an attempt is made to generate similar-sounding samples for an input. 
For this purpose a bottleneck is built into the encoder, so that the model does not simply remember the complete input.
Melody and performance conditioning synthesizes a performance with a different style for an input melody.

In \cite{zhang2020latfsmg} sequences with more than 1000 tokens are generated, based on 30 priming tokens each, which are taken from the test set for the evaluation. 

In general, Transformer architectures can generate long token sequences (\textgreater1000 tokens), which is an important characteristic for capturing the long-term structures of music.


\section{Evolutionary Computation}
Inspired by the process of natural selection \citep{Darwin1859originspeciesmeans}, genetic or evolutionary algorithms (EA) formalize music generation as an iterative cycle of recombining and mutating elements in a virtual population of musical phrases and selecting the best candidates via a fitness function for the next generation \citep{heitkotter2001hitch-hikers}. 

\citet{biles1994genjam,biles2002genjam} presented with GenJam a highly specialized system for generating traditional jazz solos based on a modified EA. In a training run, two hierarchically linked populations are generated and evaluated by interactive aesthetic selection via keyboard input. The bar-population contains one-bar melody components consisting of eighth notes and rests, which are encoded as a relative scale tone to the fundamental of the currently sounding chord. The phrase-population contains four-bar combination patterns of the bar fragments and are combined in cross-over recombination.
Furthermore, EAs can be applied in the autonomous evolution of musical synthesizers settings. To approximate a synthetically generated sound to a desired target, genetic algorithms or genetic programming can help to find the best choice of parameters for it \citep{miranda2004crossroads,horner2007evolution}.
In recent years, other parametric approaches have been increasingly integrated into EAs, e.g. autoencoders \citep{McArthur2021ApplicationEvolutionaryMusic} or GANs \citep{Marchetti2021ConvolutionalGenerativeAdversarial}.
\chapter{Co-creative Process and Evaluation}

\section{Co-creative Process}
The algorithms and models presented here exhibit varying degrees of co-creativity. Co-creativity here refers to a joint creative process of humans and machines through pre-trained knowledge and interaction in real time. 
While in compositional processes the human-machine interaction often remain in opaque curation, real-time based improvisational performances provide a more transparent insight into the underlying interaction processes.

Statistical modeling facilitates fast and steerable training mechanisms, enabling rapid adaptation during a human-machine improvisation. However, the relatively short underlying sequences do not capture a wider context and may lead to more imitative and predictable playing. 
The lack of a learning adaptation during performance is especially noticeable in formal grammars, which operate the generation of a specific grammar and its application as separate steps.
More sophisticated implementations of HMM, however, can give more precise control over the adaption rate.
This problem also applies to deep learning approaches in general. However, the accompaniment or generation of melodies based on given chords becomes applicable within a certain stylistic idiom even with pre-trained models in an interactive context.
While the ability of attantion-based models to generate longer coherent structures can lead to very convincing results, this foundational process also limits their potential for interactive scenarios. The influence of a human player on the co-creative processes of music cannot go beyond providing a primer token.

The type of representation is very crucial for the achievable level of flexibility and creative influence (Ref TISMIR paper).

\section{Evaluation}
An open problem for generative music systems remains the evaluation of their functionality and outcome. In creative tasks, by definition, there is no ground truth available for objective assessment.
It is not based on any normative aesthetics \citep{Kalonaris2021ReSoundArt} and therefore requires adapted approaches and interactive methods.

Some systems include components for \emph{internal evaluation} as part of their generative process~\citep{agres2016evaluationMusicalCreativity}, e.g., the discriminator in GANs, the fitness function in evolutionary algorithms or virtual critics trained on a specific corpus~\citep{ariza2009interrogatorCriticTuring}.  However, the quantitative aesthetic judgment is related here to a fixed framework defined by training data or algorithmic scoring and therefore only seems to serve the purpose of certain stylistic conformity.
According to \citet{colton2008CreativityPerceptionCreativity}, in order to be perceived as creative, independent methods of appreciating it's own generated artifacts are needed in addition.

For an \emph{external evaluation}, \citet{jordanous2016fourPPPPerspectivescomputational} has adapted the four different perspectives on creativity (person, process, product, press/environment) to the field of computational creativity. This widened the focus from only the generated outcomes (product) to the general capabilities of the evaluated system and its conditions. 
A common yet problematic approach to evaluating the person perspective are Turing Test style surveys. Similar to this \enquote{Imitation Game} \citep[433]{turing1950computing}, in which a human interrogator is challenged to identify the human from the machine in a virtual chat conversation, listeners are asked here to assign artifacts to human or machine creation. 
In contrast to the original conception, however, such questionnaires do not take into account the interactive aspect of the Turing test: the interrogator should perform the test in direct and improvised contact with the machine \citep{ariza2009interrogatorCriticTuring}.


Comprehensive reviews of evaluation methodologies can be found in \citet{ritchie2019evaluationCreativeSystems,agres2016evaluationMusicalCreativity}.

In \enquote{Spirio Sessions} project \citep{trump2021spirioSessions}, we combine and test different generative approaches in an interactive setting as a duo improvisation between a digital player piano and a melody instrument.
Specific criteria and guidelines for evaluating the systems involved in such interactive scenarios are the subject of further research in this project.

If no particular style of music is to be imitated, but new aesthetic positions are sought, practical studies in collaboration with musicians seem indispensable to evaluate progress in the development of interactive music systems.


\printbibliography

\end{document}